# Characterization of spatially varying aberrations for wide field-of-view microscopy

**Guoan Zheng,**[*,†] **Xiaoze Ou,**[†] **Roarke Horstmeyer, and Changhuei Yang**

*Department of Electrical Engineering, California Institute of Technology, Pasadena, CA 91125, USA*
[†]*These authors contributed equally to this work*
[*]*Correspondence: gazheng@caltech.edu*

**Abstract:** We describe a simple and robust approach for characterizing the spatially varying pupil aberrations of microscopy systems. In our demonstration with a standard microscope, we derive the location-dependent pupil transfer functions by first capturing multiple intensity images at different defocus settings. Next, a generalized pattern search algorithm is applied to recover the complex pupil functions at ~350 different spatial locations over the entire field-of-view. Parameter fitting transforms these pupil functions into accurate 2D aberration maps. We further demonstrate how these aberration maps can be applied in a phase-retrieval based microscopy setup to compensate for spatially varying aberrations and to achieve diffraction-limited performance over the entire field-of-view. We believe that this easy-to-use spatially-varying pupil characterization method may facilitate new optical imaging strategies for a variety of wide field-of-view imaging platforms.

## 1. Introduction

The characterization of optical system aberrations is critical in such applications as ophthalmology, microscopy, photolithography, and optical testing [1]. Knowledge of these different imaging platforms' aberrations allows users to predict the achievable resolution, and permits system designers to correct aberrations either actively through adaptive optics or passively with post-detection image deconvolution. Digital aberration removal techniques play an especially prominent role in computational imaging platforms aimed at achieving simple and compact optical arrangements [2]. A recent important class of such platforms is geared towards efficiently creating gigapixel images with high resolution over a wide field-of-view (FOV) [2, 3]. Given the well-known linear scaling relationship between the influence of aberrations and imaging FOV [4], it is critical to characterize their effect before camera throughput can be successfully extended to the gigapixel scale.

Over the past half-century, many unique aberration characterization methods have been reported [5-17]. Each of these methods attempts to estimate the phase deviations or the frequency response of the optical system under testing. Several relatively simple non-interferometric procedures utilize a Shack-Hartmann wavefront sensor [11-13], consisting of an array of microlenses that each focus light onto a detector. The local tilt of an incident wavefront across one microlens can be calculated from the position of its detected focal spot. Using the computed local tilts from the microlenses across the entire array, the amplitude and phase of the incident wavefront can be directly approximated. Despite offering high accuracy, measuring aberrations with a Shack-Hartmann sensor often requires considerable modification to an existing optical setup. For example, insertion and removal of the wavefront sensor from the imaging platform's pupil plane requires additional relay lenses, each subject to their own aberrations and possible misalignments.

Alternatively, wavefront aberrations can be inferred directly from intensity measurements by relying upon phase retrieval procedures [18-24]. A common phase retrieval-based strategy is to introduce phase diversity [18, 24] between multiple measurements of the intensity of an optical field. Phase diversity may be introduced either with additional optical elements or by simply inducing system defocus. Various methods for phase retrieval using defocus diversity have been reported in literature, including transport-of-intensity equation (TIE) based methods [25-28], iterative algorithms [29] and other non-iterative methods [30, 31].

By applying defocus diversity in microscopy systems, it has been shown that the complex pupil function of a high numerical aperture (NA) microscope objective lens [19, 20, 23] can be characterized with intensity-only measurements. These previous approaches, however, operated under the simplified assumption that an objective lens's aberrations do not exhibit any variation across its image plane [18-20, 23]. This approximation of a shift-invariant point-spread-function (PSF) only remains valid for objective lenses exhibiting a very small FOV. The variability of off-axis aberrations must be considered in microscopy systems or advanced computational imaging platforms that are designed to provide a very wide FOV, as their aberrated PSFs will vary significantly in shape across the image plane. These systems geared, for example, towards gigapixel photography [2, 3] and whole slide imaging [32], typically exhibit aberrations that increase as a function of distance from the image center. While prior work reports measurement of such spatially varying aberrations in lithography systems [33, 34], they unfortunately require custom designed reticle masks and interferometry setups. Precise optical alignment and specialized sample preparation are unavoidably involved, which prevents their generalized implementation within other optical pipelines.

In this paper, we describe a characterization method that is able to map spatially varying aberrations in a robust, cost-effective and easy-to-implement manner. In brief, this method operates by collecting a set of intensity images of a calibration sample at various defocus planes. The sample must contain identical discretized objects spread over its entire viewing area. In combination with a phase-retrieval algorithm, our method first recovers the phase-and-amplitude profile of a target object located at the center of the FOV. This complex profile then serves as the ground truth image of the object (i.e., image with minimal aberration). Next, our method automatically identifies another target object at an off-axis location and initializes a set of aberration parameters at that location. We then use this set of aberration parameters, in combination with the recovered ground truth image, to generate a set of aberrated intensity images for the same number of defocus planes. For each off-axis location, we recover its associated aberration parameters by minimizing the difference between the generated aberrated intensity images and the collected experimental data. Finally, we apply the recovered off-axis aberration parameters (from ~350 locations in our experiment) to generate continuous 2D aberration function maps by parameter fitting.

To demonstrate the utility of the recovered 2D aberration maps, we experimentally show how they can be used in combination with a phase-retrieval method to render images with improved resolution performance – spatially varying aberrations can be compensated by using an information-preserving image deconvolution scheme.

This paper is structured as follows: In Section 2, we briefly review some of the concepts essential to the context of our work, including phase retrieval and spatially varying pupil aberrations. In Section 3, we describe our experimental setup and the calibration sample. In Section 4, we detail our procedure for pupil function recovery at one location off the optical axis. In Section 5, we explain how to automate the aberration characterization process, experimentally demonstrate the automated measurement of spatially varying aberration weights, and show how these weights can yield accurate 2D aberration function estimates. In Section 6, we demonstrate a specific application of these aberration function maps – improving the resolution performance of phase retrieval-based image rendering across the entire imaging FOV. Finally, we end with a discussion of some of advantages and limitations of the reported method.

## 2. Overview of phase retrieval and spatially varying pupil aberrations

### 2.1 Phase retrieval and defocus diversity

The first concept essential to our work is the application of the phase retrieval algorithm using defocus diversity. As in many inverse problems, a common formulation of the phase retrieval problem is to seek a complex field solution that is consistent with measurements of its intensity. The Gerchberg-Saxton algorithm [35], as well as its related error reduction algorithm [36-38], were the first widely used numerical schemes for this type of problem. They consist of alternating enforcement of known information in the spatial and/or Fourier domains. Although phase retrieval algorithms work well for many cases of interest, stagnation and ambiguity problems are known to prevent strict convergence. A technique termed phase diversity has been developed to overcome these limitations [18, 24, 29, 39, 40]. This technique relies on measuring multiple intensity patterns with a known modification to the optical setup applied between each measurement. The set of captured images, along with the knowledge of the diversity function, is then used to iteratively converge to a complex field that agrees with each measurement. Stagnation and ambiguity problems are overcome by providing a set of measurements that more robustly constrain the phase retrieval process. Increased accuracy is guaranteed through an analysis of its Cramer-Rao lower bound [41].

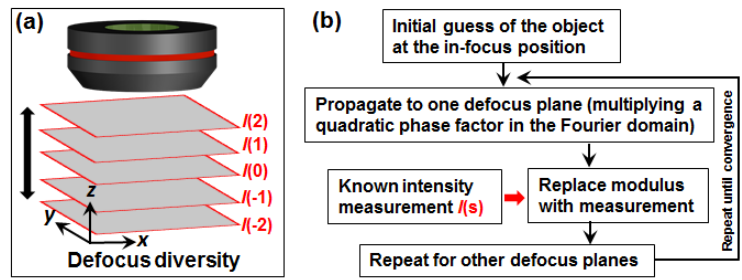

Fig. 1. Multi-plane phase retrieval with defocus diversity. (a) Multiple intensity images $I(s)$ (s = -2, -1, 0, 1, 2) are captured at different defocus settings. (b) Multi-plane iterative phase retrieval algorithm presented in [29].

In this work, we apply defocus diversity [29, 37] to perform phase retrieval within a conventional microscope. Two or more images must be captured with known defocus distances, as shown in Fig. 1(a). Based on these intensity measurements $I(s)$ (s=-2, -1, 0, 1, 2 in Fig. 1(a)) at different defocus planes, we follow the multi-plane iterative algorithm outlined in Fig. 1(b) [29]. In this algorithm, we first initialize a complex estimate of the object function. This complex estimate is then propagated to one defocus plane (multiplication by a quadratic phase factor in the Fourier domain [42]). After propagation, the amplitude of the estimate is replaced by the square root of the corresponding measurement $I$(s), while the phase is kept unchanged. Such a propagate-and-replace process is repeated until the complex solution converges (see Section 4 for implementation details).

## 2.2 Spatially varying pupil aberrations

Second, an understanding of spatially varying pupil aberrations is important to fully appreciating the impact of our work. In an aberration-free coherent imaging system, the light field distribution at the pupil plane (i.e., the back focal plane of the objective lens) is directly proportional to the Fourier transform of the light field at the object plane. Therefore, the spatial coordinates at the object plane and the pupil plane can be expressed as ($x, y$) and ($k_x, k_y$), respectively, with $k_x$ and $k_y$ as the wave number in the $x$ and $y$ directions. Due to such a Fourier relationship, aberrations of an imaging platform are often characterized at the pupil plane for simplicity [42]. Different types of aberrations can be quantified as different Zernike modes at the pupil plane. For example, defocus aberration can be modeled as a phase factor $p_5 Z_2^0(k_x, k_y)$, where $Z_2^0(k_x, k_y)$ denotes the corresponding Zernike polynomial for this aberration (here a quadratic function), while coefficient $p_5$ denotes the amount of defocus aberration (subscript '5' indicates the fifth Zernike mode).

A more complete aberration model uses the generalized pupil function $W(k_x, k_y)$, whose phase factor is a summation of different Zernike modes with different aberration coefficients $p_m$ ($p_m$ denotes the amount of $m^{th}$ Zernike mode; refer to Eq. (1) in Section 4). If the imaging platform is shift-invariant, each aberration coefficient $p_m$ is constant over the entire imaging FOV and the generalized pupil function $W(k_x, k_y)$ is independent of spatial coordinates $x$ and $y$. However, as noted above, recent extreme-FOV computational imaging platforms push beyond the limits of conventional lens design and thus invalidate this shift-invariant assumption. Aberration coefficients $p_m$s are 2D functions of x and y in this case, and thus, the generalized pupil function can be expressed as a function of both $k_x$, $k_y$ and x, y, i.e. $W(k_x, k_y, x, y)$. Our goal here is to characterize the aberration parameters $p_m$ (m=1, 2, …) as a function of spatial coordinates x and y. Based on $p_m$(x, y), we can derive the generalized pupil function $W(k_x, k_y, x, y)$ at any given spatial location (Section 5) and accurately perform post-detection image deconvolution (Section 6).

## 3. Experimental setup and sample preparation

In our experiment, we used a conventional upright microscope (BX 41, Olympus) with a 2X apochromatic lens (0.08 NA, Olympus) and a full-frame CCD camera (KAI-29050, Kodak). The tested objective lens has a relatively large FOV (~1.3 cm in diameter) with the potential to facilitate whole-slide imaging for a variety of applications [32]. However, scale-dependent geometric aberrations compound any attempt to directly capture images at a resolution commensurate with the specified NA uniformly across the entire image plane [4]. While aberrations are well-corrected near the optical axis, significant blur deteriorates image quality towards the FOV's edge.

To characterize these spatially varying aberrations of the objective lens, we first create a calibration "target" sample containing identical discretized objects over its full viewing area. While several convenient targets exist, we found that simply spin-coating a layer of 10-micron diameter microspheres (Polysciences, Inc.) on top of a microscope slide offered an ideal calibration sample. Selecting a sparse concentration of microspheres ensures that an automated search algorithm can successfully identify each microsphere, as detailed in Section 5. For example, a slide that contains 350 microspheres distributed randomly over the 1.3 cm FOV associated with the 2X objective works well.

A microsphere target sample is easy-to-prepare, cost-effective and accessible to the average microscopist. Sample preparation time totals less than 2 minutes. The standard deviation of the microspheres' size is about 0.3 μm, and thus, these calibration objects are effectively identical over the entire FOV. We note that while alternative fabrication methods such as e-beam or photo-lithography may also generate calibration samples, the aberrations of lithography lens, the evenness of photoresist, and the alignment of the mechanical stage would all need to be considered and jointly optimized to minimize unexpected target variations.

## 4. Off-axis pupil function recovery

With a proper calibration target prepared, we are now ready to detail our procedure for pupil function recovery at one location off the optical axis. Assuming the aberrations of the objective lens are minimal (i.e., they are well-corrected) at the center of its FOV, we use images of the object located near the FOV center to serve as the ground truth for other off-axis positions. The proposed characterization approach consists of two primary steps: 1) phase retrieval, and 2) pupil function estimation, as detailed below.

1) *Phase retrieval*. Following the general procedure outlined in Section 2, we displace the microscope stage from the focal plane at $\delta$ = 50 μm increments in either defocus direction, capturing a total of 17 images of the microsphere calibration target $I(s)$, where $s$ = (-8,…0,…8). The maximum defocus distance with such a scheme is 400 μm in either direction. For each image, the microsphere target is illuminated with a quasi-monochromatic collimated plane wave (632 nm).

Next, we create a $64^2$-pixel cropped image set $I_c(s)$ that contains one microsphere at the center FOV (see Fig. 2, left). We recover the complex profile of this centered microsphere using the multi-plane phase retrieval algorithm [29] from Section 2, detailed briefly as follows. First, an estimate of the complex field is initialized at the object plane. The initial estimate's phase is set to a constant and its amplitude is set to the square root of the in-focus intensity measurement of the centered microsphere $I_c(0)$. Second, this complex field estimate is Fourier transformed and multiplied by a quadratic phase factor $\exp(ik_z z)$, describing defocus of the field by axial distance $z = s \cdot \delta$. To begin, we set $s$ = 1, corresponding to $z$ = +50 μm of defocus. Third, after digitally defocusing, we again replace the amplitude values of the complex field estimate with the square root of the intensity data from recorded image, $I_c(s)$. Beginning with $s$ = 1, we first use the intensity values $I_c(1)$ captured at $z$ = +50 μm for amplitude value replacement, while the estimate's phase values remain unchanged. This digital propagate-and-replace process is repeated for all values of $s$ (all 17 cropped intensity measurements from the captured focal stack). Finally, we iterate the entire phase retrieval loop approximately 10 times. The final recovered complex image, denoted as $\sqrt{I_{truth}}\,e^{i\varphi_{truth}}$, serves as a "ground truth" estimate of the complex field from a minimally aberrated microsphere, which may be digitally refocused to any position of interest.

2) *Off-axis pupil function estimation*. Next, we select a microsphere at a position $(x_0, y_0)$ off the optical axis and generate a new $64^2$-pixel cropped image set $I_d(s)$ from our initial measurements, centered at $(x_0, y_0)$ (see Fig. 2). We also initialize an estimate of the unknown location-dependent pupil function for this position, $W(k_x, k_y, x_0, y_0)$. For simplicity, we approximate the unknown pupil function $W(k_x, k_y, x_0, y_0)$ with 8 Zernike modes, $Z_1^{-1}$, $Z_1^1$, $Z_2^{-2}$, $Z_2^2$, $Z_2^0$, $Z_3^{-1}$, $Z_3^1$ and $Z_4^0$, corresponding to x-tilt, y-tilt, x-astigmatism, y-astigmatism, defocus, x-coma, y-coma and spherical aberration, respectively [1]. The point-spread function at the selected off-axis microsphere location $(x_0, y_0)$ may be uniquely influenced by each mode above. We denote the coefficient for each Zernike mode with $p_m(x_0, y_0)$, where the subscript '$m$' stands for the mode's polynomial expansion order (in our case, $m$ = 1, 2…8). With this notation, our unknown pupil function estimate $W(k_x, k_y, x_0, y_0)$ can be expressed as,

$$W(k_x,k_y,x_0,y_0) = \exp[i\cdot\pi(p_1(x_0,y_0)Z_1^{-1}(k_x,k_y) + p_2(x_0,y_0)Z_1^1(k_x,k_y) + ...$$
$$p_3(x_0,y_0)Z_2^{-2}(k_x,k_y) + p_4(x_0,y_0)Z_2^2(k_x,k_y) + p_5(x_0,y_0)Z_2^0(k_x,k_y) + p_6(x_0,y_0)Z_3^1(k_x,k_y) + ... \quad (1)$$
$$+ p_7(x_0,y_0)Z_3^{-1}(k_x,k_y) + p_8(x_0,y_0)Z_4^0(k_x,k_y))]$$

Here, each mode $p_m(x_0, y_0)$ is a space-dependent function evaluated at $(x = x_0, y = y_0)$, allowing the pupil function $W$ to model spatially varying aberrations. This pupil function estimate is then used along with the "ground truth" complex field of the centered microsphere found in step 2 to generate a set of aberrated intensity images, $I_a(s)$, as follows:

$$I_a(s) = |\mathcal{F}^{-1}(W(k_x,k_y,x_0,y_0) \times \mathcal{F}(\sqrt{I_{truth}}e^{i\varphi_{truth}}) \times e^{ik_z\delta s}|^2, \quad (2)$$

where $\mathcal{F}$ is the Fourier transform operator and the term $e^{ik_z\delta s}$ represents defocus of the ground truth microsphere field to plane $s$. We then adjust the values of the 8 unknown Zernike coefficients $p_m$ comprising the pupil function $W$ to minimize the difference between this modeled set of aberrated intensity images $I_a(s)$ and the actual set intensity measurements of the selected off-axis microsphere, $I_d(s)$. The corresponding pupil function described by 8 Zernike coefficients is recovered when the mean-squared error difference is minimized. We apply a Generalized Pattern Search (GPS) algorithm [43] to solve the following nonlinear optimization problem for pupil function recovery:

$$(p_1, p_2 \cdots p_8)|_{(x=x_0, y=y_0)} = \underset{(p_1, p_2 \cdots p_8)}{\operatorname{argmin}} \sum_{s=-8}^{8}(\sqrt{I_a(s)} - \sqrt{I_d(s)})^2 \quad (3)$$

Based on these optimal Zernike coefficients, the off-axis pupil function can be approximated following Eq. (1). Determining the aberration function associated with one off-axis microsphere requires an approximate computation time of 90 seconds on a personal computer with an Intel i7 CPU.

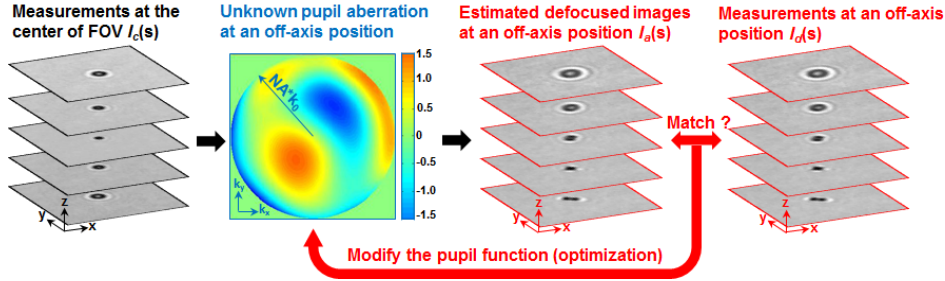

Fig. 2. Pupil function recovery at one off-axis position. Two cropped areas of one set of defocused intensity images are used for algorithm input. One cropped set $I_c(s)$ is centered on a microsphere at the images' central FOV (left), while the other cropped set $I_d(s)$ is centered on a microsphere at an off-axis position (right). Each cropped image set contains 17 intensity measurements (here only 5 are shown) at different defocus distances (-400 μm to +400 μm, 50 μm per step). We approximate an unknown pupil function $W$ with 8 Zernike coefficients (x-tilt, y-tilt, x-astigmatism, y-astigmatism, defocus, x-coma, y-coma and spherical aberration). We use this pupil function estimate to modify the 17 "ground truth" images $I_c(s)$ of the central microsphere to generate a new set of aberrated intensity images, $I_a(s)$ (middle). We then adjust the values of the 8 unknown Zernike coefficients to minimize the difference between $I_a(s)$ and the actual intensity measurements of the off-axis microsphere, $I_d(s)$ (right). The corresponding pupil function described by 8 Zernike coefficients is recovered when the mean-squared error difference between these two image sets is minimized.

## 5. Spatially varying aberration characterization over the entire FOV

Repeating the previous section's off-axis aberration recovery scheme for many different microspheres spread over the image plane, we are able to characterize a microscope objective's spatially varying aberrations over its entire FOV. The center of each microsphere is automatically identified using a marker-controlled watershed segmentation algorithm [44]. We also measure the distance between each marked microsphere and its nearest neighbor.

Any microsphere within a 150 μm radius of a neighbor is automatically skipped to avoid multiple computations at sphere clusters.

Fig. 3(a) shows a full FOV image of the calibration target with ~350 microspheres denoted by a red dot. For each microsphere, we recover the same 8 location-specific Zernike coefficients. For example, Fig. 3(b) shows the pupil function $W$ recovered following Eq. (3) at position $(x_1, y_1)$, the center of the black square in Fig. 3(a). Fig. 3(c1)-(c5) are 5 of the 17 intensity measurements of the microsphere at position $(x_1, y_1)$ under different amounts of defocus: $I_d(s = 0)$, $I_d(s = \pm 3)$, and $I_d(s = \pm 6)$. Fig. 3(d1)-(d5) display the corresponding aberrated image estimates $I_a(s)$ generated by the recovered pupil function in Fig. 3(b). Following the convex form of Eq. (3), the applied GPS algorithm successfully minimizes the mean-squared error difference between the measurements $I_d(s)$ and the estimates $I_a(s)$.

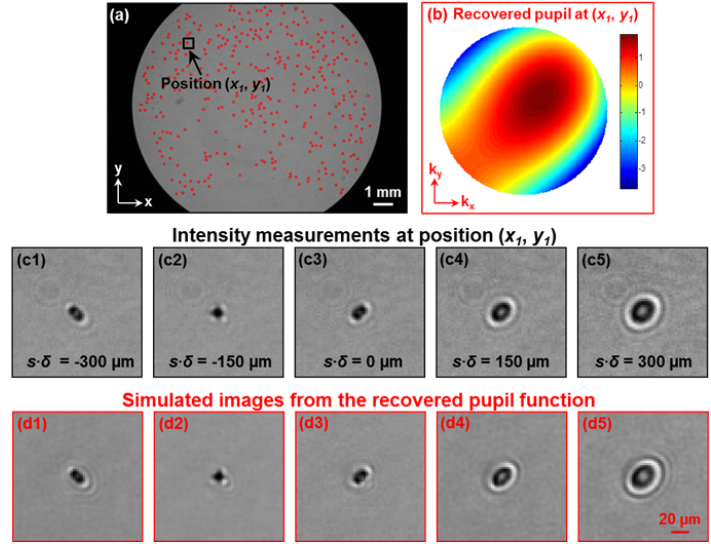

Fig. 3. Off-axis aberration characterization with a calibration target. (a) ~350 microspheres are automatically identified on a microscope slide, each denoted by a red dot. (b) The recovered pupil function at position $(x_1, y_1)$. (c1)-(c5) Intensity measurements $I_d(s)$ of the microsphere centered at $(x_1, y_1)$ under different amounts of defocus. (d1)-(d5) The corresponding aberrated image estimates generated using the pupil function in Fig. 3(b).

Following this aberration recovery pipeline, 8 Zernike coefficients are calculated for approximately 350 unique spatial locations across the microscope's FOV. Fig. 4(a)-(f) plot the recovered second, third and fourth order spatially varying aberrations of our tested 2X objective lens, corresponding to x-astigmatism, y-astigmatism, defocus, x-coma, y-coma and spherical aberration respectively (first order Zernike modes are normally not considered as aberrations, and are thus not shown). The full FOV image of our calibration target is displayed at the bottom plane of each plot, where the FOV diameter is 1.3 cm. Each blue dot in Fig. 4 represents the recovered coefficient for the corresponding Zernike mode, and the spatial location of each blue dot corresponds to one microsphere labeled in Fig. 3(a).

Finally, we fit these 350 discrete values to a continuous polynomial function $p_m(x, y)$, allowing us to accurately recover the pupil function at any location across the image plane (curved surfaces in Fig. 4). The order of each polynomial function can be predicted via aberration theory for a conventional imaging platform [1]. The aberrations of increasingly unconventional optical designs in computational imaging systems may not follow such predictable trends, which we may account for with alternative fitting models and/or recovering coefficients at more than 350 unique spatial locations.

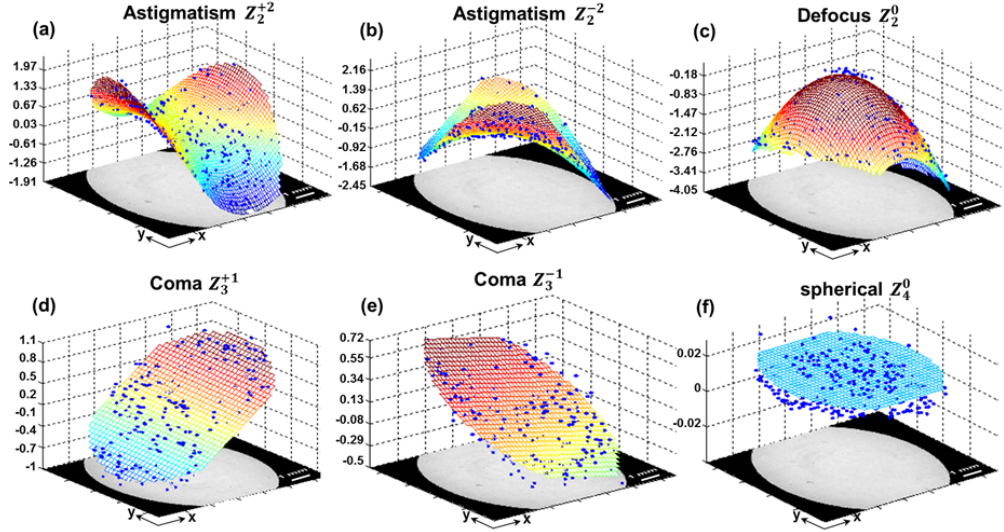

Fig. 4 Spatially varying aberrations of the 2X objective lens. Each data point, denoted by a blue dot, represents the extracted Zernike coefficient weight for one microsphere. ~350 microspheres are identified over the entire FOV and their corresponding parameters are fitted to a 2D surface for each type of aberration. (a)-(f) correspond to x-astigmatism, y-astigmatism defocus, x-coma, y-coma and spherical aberration.

We verified the accuracy of our aberration parameter recovery process with an additional simple experiment. We defocused the calibration target by +50 µm along the optical axis and again implemented our aberration parameter recovery process (using the same ground truth images as before). For the tested wide-field microscope objective, Fig. 5 displays two of these fitted polynomial functions for spatially varying defocus - one computed for an in-focus target and one for the target under +50 µm of defocus. The major difference between the two polynomial fits is a constant offset corresponding to Δz = 48.9 µm, which is in a good agreement with the experimentally induced +50 µm displacement distance. As a reference, the depth-of-focus of the objective lens is about 80 µm.

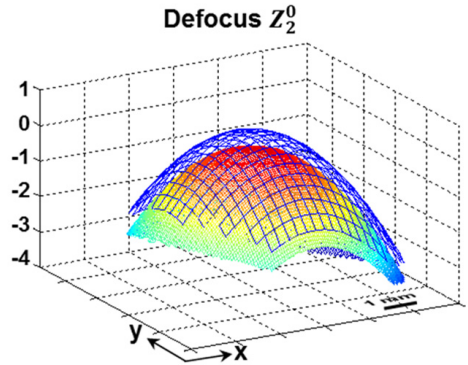

Fig. 5 Recovered defocus parameter function $p_5(x, y)$ with (color surface) and without (blue grid) +50 µm of sample defocus. The difference between these two surfaces corresponds to a defocus distance of +48.9 µm, which is in a good agreement with the actual displacement distance.

## 6. Image deconvolution using the recovered aberration parameters

We will now demonstrate that our recovered 2D aberration maps can be used in an image deconvolution process to render images with improved resolution performance. The image deconvolution process is comprised of two main steps: 1) phase retrieval, 2) segment decomposition and shift-invariant image deconvolution, as outlined below.

1) *Full-FOV phase retrieval.* We use the multi-plane phase retrieval algorithm described in Section 2 to recover the amplitude and phase of a sample over the microscope's entire FOV. This complex image contains the objective lens's spatially varying aberrations.

2) *Segment decomposition and shift-invariant image deconvolution.* We then divide the full-FOV complex image into smaller 128 x 128 pixel image segments, denoted by $I_{seg}(n)$ ($n$ = 1, 2,... 1600 for our employed detector). Aberrations within each small segment are treated as shift-invariant, a common strategy for wide FOV imaging processing [45]. The pupil function $W(k_x, k_y, x_c(n), y_c(n))$ is then calculated for each small segment following Eq. (1), where ($x_c(n)$, $y_c(n)$) represents the central spatial location of the $n^{th}$ segment. We then perform image deconvolution to recover the corrected image segment $I_{cor}(n)$ as follows:

$$I_{cor}(n) = \left| \mathcal{F}^{-1}(\mathcal{F}(\sqrt{I_{seg}}\, e^{i\varphi_{seg}}) / W(k_x, k_y, x_c(n), y(n)_c)) \right|^2, \quad (4)$$

where $\sqrt{I_{seg}}\, e^{i\varphi_{seg}}$ is the corresponding cropped segment of the complex field recovered in Step 1. We note that, in the above equation, we only perform division within the circular pupil of the objective lens; for regions outside the circular pupil, we set the spectrum to 0 in the Fourier domain. Furthermore, since our deconvolution process is applied to complex data, we successfully avoid division by zero in the Fourier domain.

To characterize the resolution performance of the above deconvolution process at different image plane locations, we perform a first experiment using a shifted USAF resolution target as our sample. Fig. 6(b1)-(d1) are the raw image segments $I_{seg}$ directly captured using the aberrated objective lens, while Fig. 6(b2)-(d2) are the corresponding processed images $I_{cor}$ using Eq. (4). From Fig. 6(b2)-(d2), Group 7, element 1 (line width of 3.9 μm) of the USAF target can be resolved, in a good agreement with the Abbe diffraction limit of 3.94 μm of our 0.08 NA objective lens. This simple experiment indicates our aberration correction scheme can correct this particular objective's aberration blur to yield diffraction-limited performance across its entire image FOV.

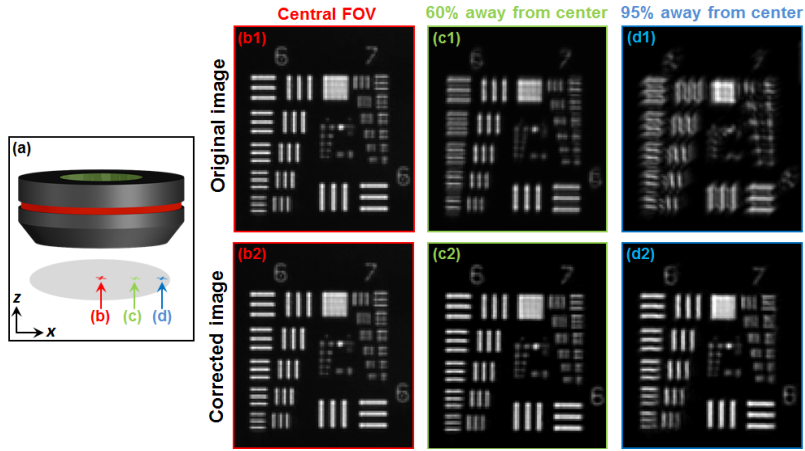

Fig. 6 Resolution characterization using a USAF resolution target. (a) The USAF resolution target is placed at 3 different locations indicated by color arrows (b)-(d). Full FOV corresponds to circular region with 1.3 cm diameter. The original images captured using the aberrated objective lens at the center (b1), 60% away from the center (c1), and 95% away from the center (d1). (b2)-(d2) are the corresponding processed images using the deconvolution scheme. Group 7, element 1 (line width of 3.9 μm) of the USAF target can be resolved from the corrected images, in a good agreement with the Abbe diffraction limit of 3.94 μm.

Based on Eq. (4), we can also recombine all the corrected image segments $I_{cor}(n)$ to form a correct full FOV image. Fig. 7 and Fig. 8 show the results of a second experiment, where the entire FOV of images of two samples are corrected. An alpha blending algorithm [46] is used to remove edge artifacts at the segment boundary. Specifically, we cut away 2 pixels at the edge of each segment and use another 5 pixels to overlap with the adjacent portions. This blending comes at a small computational cost of redundantly processing the regions of overlap twice.

The sample in Fig. 7 is the calibration target discussed in Section 3, and the sample in Fig. 8 is a new test target with a mixture of microspheres of different diameters (5-20 μm) on a microscope slide. The 4 regions outlined by red squares in Fig. 7(a) and Fig. 8(a) are highlighted for detailed observation. The corresponding pupil functions of these four regions are shown in Fig. 7(b1)-(e1) and Fig. 8(b1)-(e1). Fig. 7(b2)-(e2) and Fig. 8(b2)-(e2) display their associated corrected (i.e., deconvolved) images, while Fig. 7(b3)-(e3) and Fig. 8(b3)-(e3) display their original images without aberration correction. From these two examples, it is clear that our aberration characterization procedure can digitally compensate for the spatially varying aberrations across a microscope objective's full FOV.

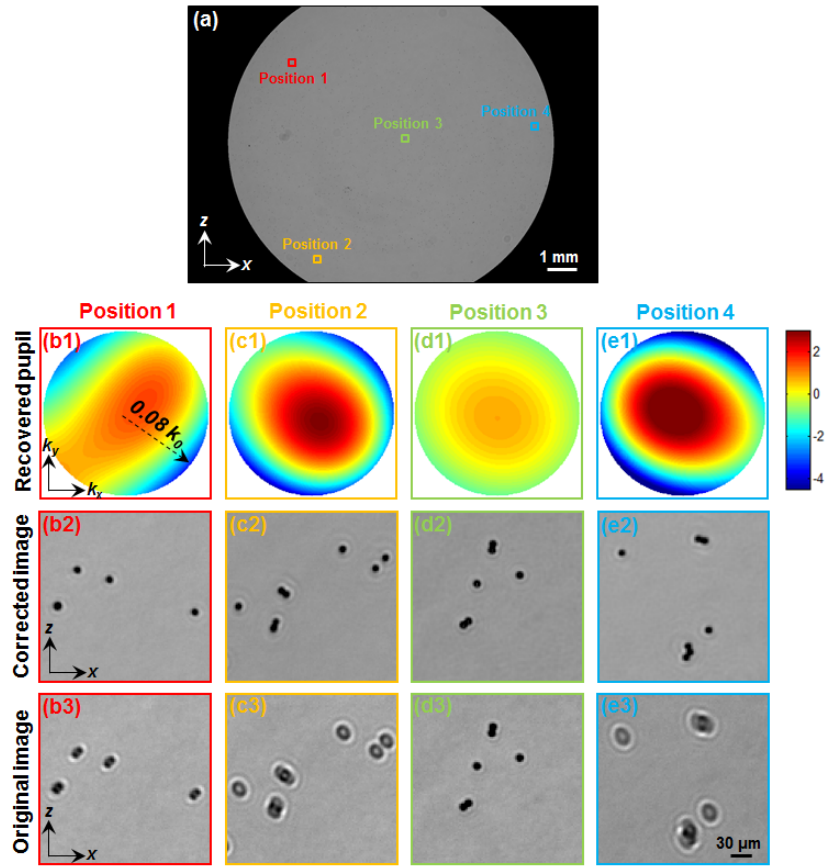

Fig. 7 Full FOV image deconvolution of the microsphere calibration target. (a) The aberration-corrected full FOV image. (b1)-(e1) Recovered pupil functions corresponding to highlighted regions in (a). (b2)-(e2) The corrected images of highlighted regions in (a). (b3)-(e3) The original images of the test target without aberration correction.

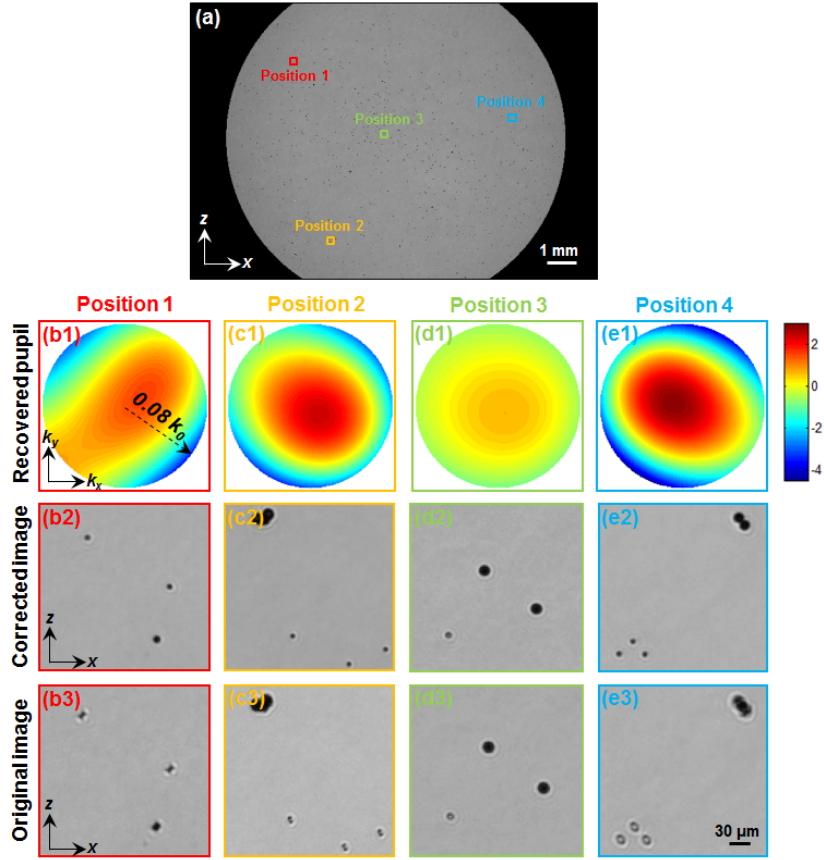

Fig. 8 Full FOV image deconvolution of a new test target, containing a mixture of microspheres with different diameters (5-20 µm). (a) The aberration-corrected full FOV image. (b1)-(e1) Recovered pupil functions corresponding to highlighted regions in (a). (b2)-(e2) The corrected images of highlighted regions in (a). (b3)-(e3) The original images of the test target without aberration correction.

Finally, we note that the deconvolution scheme in Eq. (4) is based on inverting the coherent transfer function (i.e., the complex pupil function) of the objective lens. For the case of incoherent illumination, the incoherent optical transfer function can be directly calculated from the complex pupil function through a close form equation [42], and image deconvolution can be performed in the Fourier domain accordingly.

## 7. Conclusion

In summary, we report a phase retrieval-based procedure to efficiently recover the spatially varying wavefront aberrations common in wide-FOV imaging systems. In our demonstration, we applied a generalized pattern search algorithm to measure the spatially varying aberration coefficients of a wide-FOV microscope objective at ~350 off-axis positions. These pupil functions were then used to generate 2D aberration maps by parameter fitting. We demonstrated the application of our characterization process with an example of shift-variant image deconvolution, which successfully accounts for induced aberrations over a 2X objective's entire FOV (1.3 cm diameter). The proposed computational approach does not require any optical modifications or additional hardware. The entire aberration recovery process is fully automated and easy to implement. We believe the characterization of spatially varying pupil aberrations is an attractive way to quantify the performance of many wide-FOV image platforms.

In our characterization scheme, we assume that the aberration is well-corrected at the center FOV. The object located at the center FOV serves as the ground truth for off-axis positions. If the objective lens under testing is not well-corrected at the center of the FOV, we can use other well-corrected optics (such as a high NA, small FOV objective) to capture the ground truth image. Finally, we note that future work will be aimed at extending the proposed aberration characterization pipeline beyond recovery of 8 Zernike modes. For more unconventional imaging designs, 10-15 Zernike modes may be required for accurate aberration characterization. A GPU implementation of the proposed pipeline can significantly shorten the associated processing time. Furthermore, this work tested an objective lens with an assumed 100% transmissive circular back aperture. Using our framework to model objective lens apertures with non-perfect transmission, or containing apodizing filters or coded modulation masks will be an additional future research direction.

**Acknowledgements**

We acknowledge funding support from National Institute of Health under Grant No. 1R01AI096226-01.